\documentclass[10pt]{article}
\usepackage{graphicx}
\begin{document}
\title{\bf A TBA description of thermal transport\\ in the XXZ
Heisenberg model}
\author{X. Zotos\\
Department of Physics,
University of Crete,\\71003 Heraklion, Greece\\
Max-Planck-Institut f\"ur Physik Komplexer Systeme,\\
N\"othnitzer Strasse 38, 01187 Dresden, Germany}
\date{}
\maketitle

\begin{abstract}
It is shown that the Bethe ansatz formulation 
of the easy-plane 1D Heisenberg model thermodynamics (TBA) 
by Takahashi and Suzuki \cite{ts} and the 
subsequent analysis of the spin Drude weight \cite{xz}, also reproduces 
the thermal Drude weight and magnetothermal coefficient obtained 
by the Quantum Transfer Matrix method (QTM) \cite{klumper}.
It can also be extended to study the far-out of equilibrium energy 
current generated at the interface between two semi-infinite chains held at 
different temperatures. 
\end{abstract}

The one dimensional spin-1/2 Heisenberg is the prototype model 
integrable by the Bethe ansatz method \cite{bethe,baxter,faddeev}.
Its thermodynamic properties have been studied, first by 
a method proposed by Takahashi and Suzuki (TS) \cite{ts} along the lines 
proposed by Yang and Yang \cite{yy} and later 
by a transfer matrix nethod (QTM) proposed by Kl\"umper \cite{klumper}. 

Concerning the spin and thermal transport,  
the existence of  conservation laws generically implies 
unconventional - ballistic - transport \cite{znp}. In particular, 
the energy current commutes with the Hamiltonian resulting 
to purely ballistic thermal transport characterized 
by the thermal Drude weight $D_{th}$. This fact has further 
promoted the study of thermal conduction by magnetic excitations in novel,
high quality, quasi-one dimensional magnetic compounds \cite{hess}.
The temperature and magnetic field dependence of $D_{th}$ has been evaluated 
by an extension of the QTM method \cite{ks,sk}.

On the spin transport, the situation is more involved as the spin 
current is not a conserved quantity. Although ballistic transport 
can be established at finite magnetization using the 
Mazur inequality \cite{znp},
at zero magnetization a finite spin Drude weight was found at all temperatures 
in the easy-plane regime using a TBA approach \cite{xz}.
The results of this  approach were recently corroborated by the 
finding of a quasi-local conervation law that provides a bound on the spin 
Drude weight \cite{prosen,affleck}. 
In this note, it is pointed out that the thermal Drude weight $D_{th}$ 
can similarly be obtained by the formulation 
of TS. The analysis of the spin Drude weight \cite{xz} is closely followed.

Concerning the far-out of equilibrium thermal conductance, 
the idea that the steady state energy current between two semi-infinite 
ballistic systems held at different temperatures is given by 
a Landauer description has a long history; it has   
recently been investigated in a series of pioneering studies 
\cite{erdos,doyon,doyonsg,moore,viti}. 
Here it is shown that recent numerical simulation results of the 
non-equilibrium steady current can be fairly accurately reproduced by 
the same TBA analysis.

The proposed fermionic quasi-particle description of the thermal Drude weight 
and far-out of equilibrium thermal current is rather appealing as 
it promotes a semi-phenomenological description of this system.

\bigskip
The XXZ anisotropic Heisenberg Hamiltonian for a chain of $N$ sites
with periodic boundary conditions $\sigma_{N+1}^a=\sigma_1^a$ is given by,

\begin{equation}
H=J \sum_{i=1}^N(\frac{1}{2} e^{i\phi} \sigma_i^+ \sigma_{i+1}^-+ h.c.) 
+\frac{\Delta}{4} \sigma_i^z \sigma_{i+1}^z -\frac{h}{2}\sigma_i^z,
\label{heis}
\end{equation}
where $\sigma_i^a$ are Pauli spin operators and a spin current generating 
fictitious flux $\phi$ is introduced \cite{kohn,znp}.
The region $0\leq \Delta \leq 1$ is commony 
parametrized by $\Delta=\cos\theta$, 
$J$ is taken as the unit of energy. 
For completness, a concise description of the TS formulation follows.

The pseudomomenta $k_\alpha$ characterizing the Bethe ansatz
wavefunctions are expressed in terms of the rapidities $x_\alpha$,

\begin{equation}
\cot(\frac{k_{\alpha}}{2})=\cot(\frac{\theta}{2})
\tanh(\frac{\theta x_{\alpha}}{2}).
\label{param}
\end{equation}
For M down spins and N-M up spins the energy E and momentum K are given by:
\begin{equation}
E=J\sum_{\alpha=1}^M (\cos k_{\alpha}-\Delta),~~~~
K=\sum_{\alpha=1}^M k_{\alpha}.
\label{enem}
\end{equation}
Imposing periodic boundary conditions on the Bethe ansatz wavefunctions
the following relations hold,
\begin{equation}
{\Big\lbrace} \frac{\sinh \frac{1}{2}\theta(x_\alpha+i)}
{\sinh\frac{1}{2}\theta(x_\alpha-i)}{\Big\rbrace}^N
=-e^{i\phi N} \prod_{\beta=1}^M
{\Big\lbrace} \frac{\sinh \frac{1}{2}\theta(x_\alpha-x_\beta+2i)}
{\sinh\frac{1}{2}\theta(x_\alpha-x_\beta-2i)}{\Big\rbrace};
~~~\alpha=1,2,...M.
\label{pbce}
\end{equation}
In the thermodynamic limit, the solutions of equations (\ref{pbce}) are 
grouped into
strings of order $n_j, j=1,...,\nu$ and parity $v_j=+~ {\rm or}~ -$. 
Hereafter, for simplicity, $\theta$ is limited to 
the case $\theta=\pi/\nu$ where the allowed
strings are of order $n_j=j, j=1,...,\nu-1$ and parity $v_j=+$ of the form,
\begin{equation}
x_{\alpha,+}^{n,k}=x_{\alpha}^n+(n+1-2k)i+O(e^{-\delta N});~~~k=1,2,...n,
\label{s1}
\end{equation}
and strings of order $n_{\nu}=1$ and parity $v_{\nu}=-$ of the form,
\begin{equation}
x_{\alpha,-}=x_{\alpha}+i\nu+O(e^{-\delta N}),~~~\delta > 0. 
\label{s2}
\end{equation}

In the thermodynamic limit the densities of excitations $\rho_j$ and 
holes $\rho_j^h$ are given by, 
\begin{equation}
a_j=\lambda_j(\rho_j+\rho_j^h)+\sum_k T_{jk}\ast \rho_k,~~~
a_j(x)=\frac{v_j}{2\nu} \frac{ \sin( \frac{n_j\pi}{\nu} ) }
{ \cosh( \frac{\pi x}{\nu} )-v_j \cos( \frac{n_j\pi}{\nu}) }
\label{alpha}
\end{equation}
\noindent
with $a\ast b(x)=\int_{-\infty}^{+\infty} a(x-y)b(y)dy$ and $T_{jk}$ 
the phase shifts given by TS.
The sum over $k$ is constrained to the allowed strings, given in our case by  
the equations (\ref{s1},\ref{s2}) and $\lambda_j=1, j=1,...,\nu-1$,  
$\lambda_{\nu}=-1$

To study thermal transport, it is noted that the derivatives of the
monodromy matrix with respect to the spectral parameter generate the 
conservation laws of the system \cite{faddeev}. 
In particular, the first derivative 
generates the Hamiltonian $H$, while the second the energy current $J_E$.
The energy and entropy of a BA state are given by,
\begin{equation}
E/N=\sum_j\int_{-\infty}^{+\infty}dx~
(\epsilon_j^{(0)}(\phi)-hn_j + \xi_j j_j^{(0)} ) \rho_j
\label{energy}
\end{equation}
\begin{equation}
S/N=\sum_j\int_{-\infty}^{+\infty}dx~
(\rho_j+\rho_j^h)\ln (\rho_j+\rho_j^h) -\rho_j
\ln \rho_j - \rho_j^h\ln \rho_j^h
\label{enropy}
\end{equation}
where $\epsilon^{(0)}_j=-A a_j~(A=2\nu\sin(\pi/\nu)J)$ are the zero 
excitation energies and
$j_j^{(0)}=A\partial \epsilon_j^{(0)}/\partial x$
the corresponding eigenvalues of the energy current operator. 
The fictitious fields $\xi_j$ coupled to the eigenvalues 
of the conserved energy current operator $J_E$ are introduced. 
Minimizing the free energy the standard Bethe ansatz equations for 
the  temperature dependent effective dispersions are given by,
$\epsilon_j$ at temperature $T$,
\begin{equation}
\epsilon_j=( \epsilon^{(0)}_j(\phi) - hn_j+\xi_j j_j^{(0)} ) 
+T\sum_k \lambda_ kT_{jk}\ast 
\ln(1+e^{-\beta\epsilon_k}).
\label{effe}
\end{equation}

In \cite{xz} the spin Drude weight was obtained in the form,
\begin{equation}
D=\frac{\beta}{2}\sum_j\int_{-\infty}^{+\infty}dx (\rho_j+\rho_j^h)
<n_j>(1-<n_j>) ({j_j^s}^2) 
\label{ds}
\end{equation}
with $<n_j>=1/(1+e^{\beta\epsilon_j})$ and 
$j_j^s$, related to the phase dependence $\phi$ of the energy eigenvalues,  
can be interpreted as the effective spin current of the string excitations 
given by the solution of the corresponding 
BA non-linear integral equations.
It suggested an interesting interpretation by comparing it to the analogous  
expression for independent fermions.

To derive an analogous expresion for the thermal Drude weight 
the free energy density $f$ can first be written in the convenient form,
\begin{equation}
f=F/N=-T\sum_j\int_{-\infty}^{+\infty} dx 
\lambda_j a_j \ln (1+e^{-\beta\epsilon_j}).
\label{free}
\end{equation}
The thermal Drude weight is given by the 2nd derivative with respect 
to $\xi$ ($\xi_j(x)=\xi$),
\begin{equation}
D_{th}=\frac{\beta^2}{2N} <J_E^2>,
~~~-T\frac{\partial^2 f}{\partial \xi^2}{\Big |}_{\xi=0} =<J_E^2>
\label{dth}
\end{equation}
where $J_E$ is the energy current operator \cite{znp}.
Using the key observation 
$\partial \epsilon_j /\partial \xi = A\partial \epsilon_j/\partial x$ 
and following manipulations in \cite{yy} (multiplying (\ref{effe}) by 
$\rho_j$, integrating over $x$ and summing over $j$),  
\begin{equation}
<J_E^2>/N=\sum_j\int_{-\infty}^{+\infty}dx (\rho_j+\rho_j^h)
<n_j>(1-<n_j>) ({j_j^{\epsilon}}^2)
\label{jeje}
\end{equation}
with $j_j^{\epsilon}=A\partial \epsilon_j / \partial x$ 
the effective energy current of the string excitations. 
This formulation reproduces the data of the QTM \cite{ks} 
as a function of temperature and magnetic field for $\Delta=\cos(\pi/\nu)$.

Along the same line the magnetothermal correlation  
can be evaluated by,
\begin{eqnarray}
<J_S J_E>/N&=&-T\frac{\partial^2 f}
{\partial \phi \partial \xi}{\Big|}_{\phi,\xi=0}\nonumber\\
&=& \sum_j\int_{-\infty}^{+\infty}dx (\rho_j+\rho_j^h)
<n_j>(1-<n_j>) (j_j^s j_j^{\epsilon}).
\label{jejs}
\end{eqnarray}
Note that the coincidence of $<J_EJ_S>$ results with the corresponding 
QTM data \cite{sk}, provides another view 
on the "rigid string hypothesis" \cite{fk, xz}.
While the QTM is elegant and powerful as it provides 
straightforward data at all values of the anisotropy parameter $\Delta$, 
the TBA approach offers an appealing fermionic quasi-particle 
picture. 

In recent studies of far-out equilibrium thermal transport, 
two semi-infinite chains (L - left, R - right)  
initially held at different temperatures $T_L, T_R$ are brought into 
contact and the  long time energy current at the junction is observed.
A physically motivated ansatz to evaluate the expectation value of the 
current $<J_E>$ at the junction 
is to assume that the far from the junction thermal baths 
impose left (right) currents $<J_{L(R)}>$ so that $<J_E>=<J_L> + <J_R>$.  
$<J_{L(R)}>$ is the energy current carried by positive (negative) velocity 
excitations.
Of course this ansatz relies on the integrability of the XXZ chain 
implying that each eigenstate is also an eigenstate of 
the energy current operator.  Using the relation
\begin{equation}
<J_L>=\frac{\partial f}{\partial \xi_L}{\Big |}_{\xi_L=0},
\label{jl}
\end{equation}
with the condition $\xi_j(x)=\xi_L$ for $x > 0$ and $\xi_j(x)=0$ for $x < 0$
at $\beta_L=1/T_L$, $<J_L>$ is given by, 
\begin{equation}
<J_L>=\sum_j\int_{0}^{+\infty}dx \lambda_j a_j
<n_j> j_j^{\epsilon}.
\label{expjl}
\end{equation}
$j_j^{\epsilon}=\partial \epsilon_j /\partial \xi$ is 
obtained from (\ref{effe}) under the above condition by,

\begin{eqnarray}
\frac{\partial \epsilon_j}{\partial \xi}{\Big |}_{x>0}&=&j_j^{(0)}  
-\sum_k \lambda_ kT_{jk}\ast ( <n_k> \frac{\epsilon_k}{\partial \xi})
\nonumber\\
\frac{\partial \epsilon_j}{\partial \xi}{\Big |}_{x<0}&=&  
-\sum_k \lambda_ kT_{jk}\ast ( <n_k> \frac{\epsilon_k}{\partial \xi}).
\label{jel}
\end{eqnarray}

\begin{figure}[!ht]
\begin{center}
\includegraphics[width=7.0cm, angle=-90]{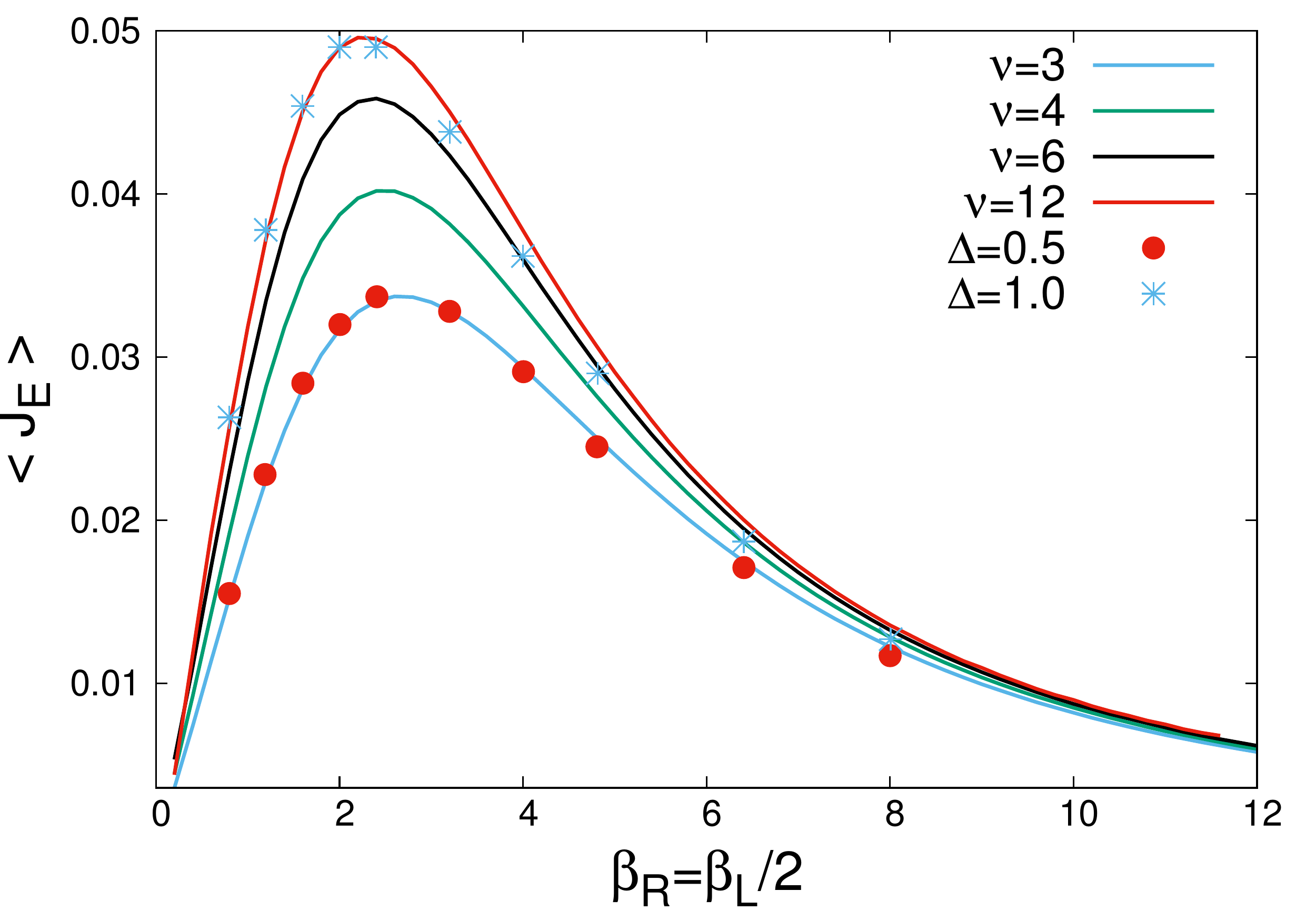}
\caption{Energy current as a function of temperature difference under the 
condition $\beta_R=\beta_L/2$; symbols are  
DMRG data from \cite{viti}.}
\label{blr}
\end{center}
\end{figure}

\begin{figure}[!ht]
\begin{center}
\includegraphics[width=7.0cm, angle=-90]{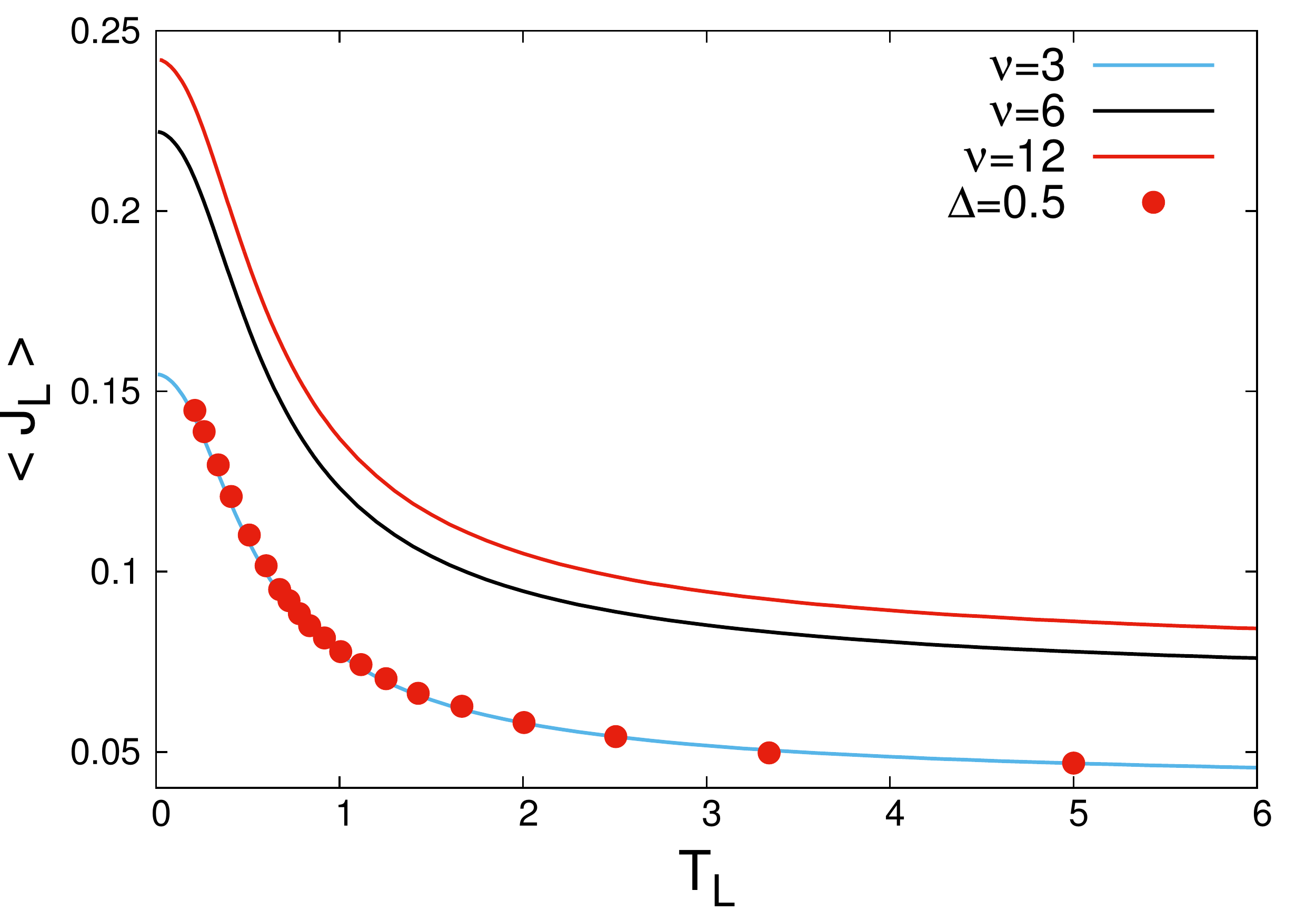}
\caption{Energy current as a function of temperature; symbols are DMRG data 
from \cite{moore} shifted by $< J_L (T_L = \infty)>$.}
\label{tr}
\end{center}
\end{figure}

Symmetrically, $<J_R>$ (of opposite sign) 
is evaluated at inverse temperature $\beta_R=1/T_R$ 
by integrating over $-\infty < x < 0$.
At the moment, the validity of this ansatz only rests on its fair agreement 
with recent DMRG numerical simulation studies \cite{moore,viti} 
as shown in Figures \ref{blr} and \ref{tr}.
It has been demonstrated in the framework of CFT's \cite{doyon} and questioned 
in the context of integrable field theories \cite{doyonsg}.
Note that in the zero temperature limit $<J_L>$ behaves 
as expected $<J_L>=(\pi/12)T_L^2$. 

\begin{figure}[!ht]
\begin{center}
\includegraphics[width=7.0cm, angle=-90]{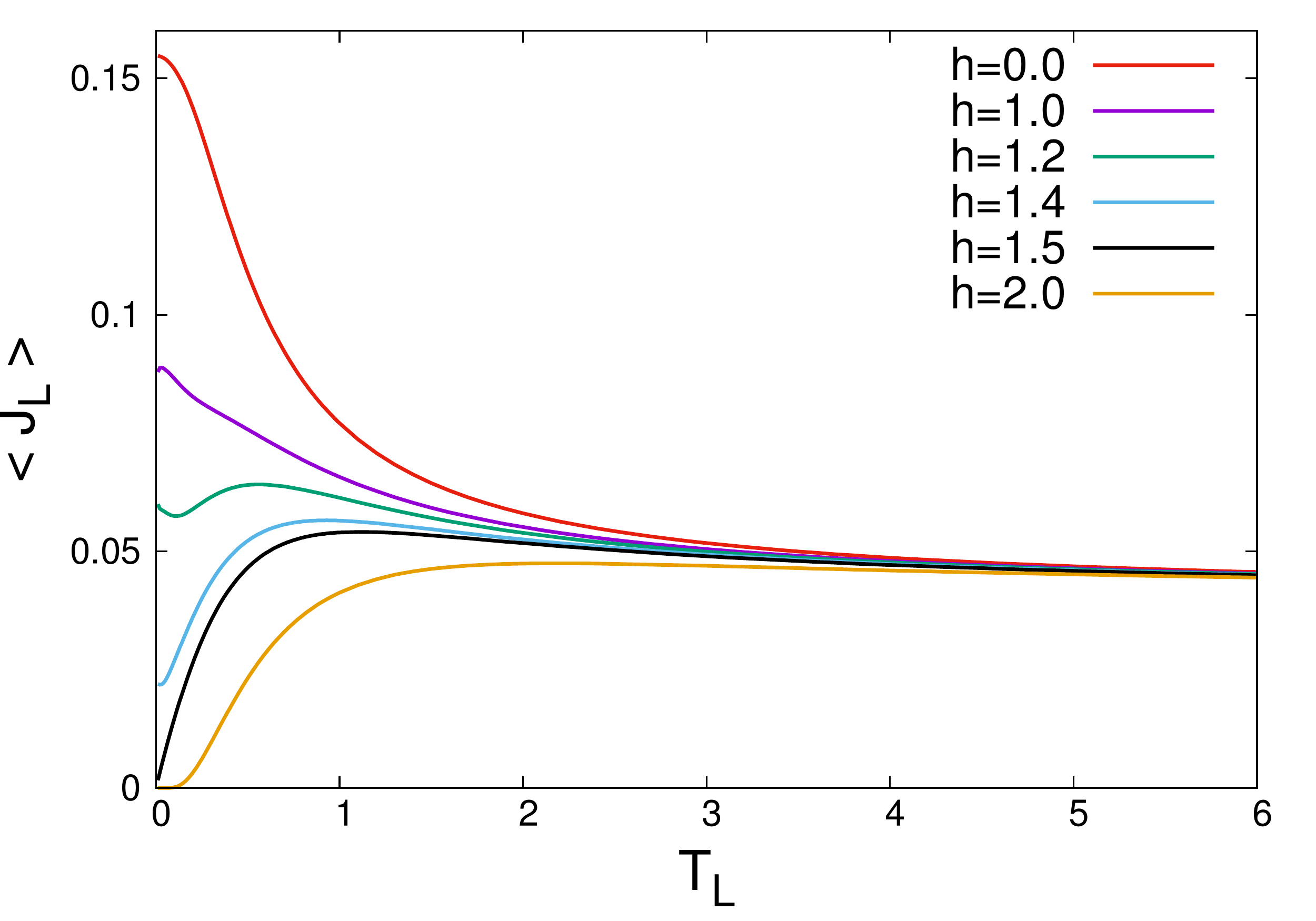}
\caption{Energy current as a function of temperature at $\Delta=0.5$ 
for different magnetic fields.}
\label{jh}
\end{center}
\end{figure}

Finally, the dependence of the current at different magnetic fields 
is presented in Figure \ref{jh}, where the critical field from the 
gapless antiferromagnetic to the gapped ferromagnetic state occurs 
at $h=J(1+\Delta)=1.5$. The non-monotonic behavior observed approaching the 
critical field is puzzling but reminiscent of a similar behavior 
of other magnetothermal quantities \cite{psaroudaki}. 
Of course in a finite magnetic field the thermal current has also a component 
from the spin current, that although ballistic, is not conserved.

It is interesting to study the applicability of this formulation 
to baths in a magnetic field, at different anisotropy parameters,
other integrable systems on a lattice with conserved currents as well as 
integrable field theories.

This work was supported by the European Union Program
No. FP7-REGPOT-2012-2013-1 under Grant No.~316165.


\begin{thebibliography}{99}
\bibitem{ts} M. Takahashi and M. Suzuki, Prog. Theor. Phys.
{\bf 48}, 2187 (1972).
\bibitem{xz} X. Zotos, Phys. Rev. Lett. {\bf 82}, 1764 (1999).
\bibitem{klumper} 
A. Kl\"umper: Z. Phys. B{\bf 91}, 507 (1993).
\bibitem{bethe} 
H. Bethe, Z. Phys. {\bf 71}, 205 (1931).
\bibitem{baxter} 
R.J. Baxter, Ann. of Phys. {\bf 70}, 193,323 (1972).
\bibitem{faddeev} 
L.Faddeev, Sov. Sci. Reviews, Harwood Academic, London, 
{\bf C1}, 107 (1980).
\bibitem{znp} X. Zotos, F. Naef and P. Prelov\v sek, 
Phys. Rev. B{\bf 55}, 11029 (1997).
\bibitem{yy} 
C.N. Yang and C.P. Yang, J. Math. Phys. {\bf 10}, 1115 (1969).
\bibitem{hess} C. Hess, Eur. Phys. J. Special Topics {\bf 151}, 73 (2007).
\bibitem{ks} A. Kl\"{u}mper and K.~Sakai, 
J. Phys. A {\bf 35}, 2173 (2002).
\bibitem{sk} K. Sakai, and A. Kl\"{u}mper, 
J. Phys. Soc. Jpn. Suppl. {\bf 74}, 196 (2005).
\bibitem{prosen} T. Prosen, Phys. Rev. Lett. {\bf 106}, 217206 (2011)
\bibitem{affleck} R.G. Pereira, V. Pasquier, J. Sirker and I. Affleck, 
J. Stat. Mech. P09037  (2014).
\bibitem{erdos} P. Erd\"os, Phys. Rev. {\bf 139}, A1249 (1965). 
\bibitem{doyon} D. Bernard and B. Doyon, J. Phys. Math. Theor. {\bf 45}, 
362001 (2012); review article in JSTAT; arXiv:1603.07765
\bibitem{doyonsg} O. Castro-Alvaredo, Y. Chen, B. Doyon and M. Hoogeveen, 
 J. Stat. Mech. P03011 (2014).
\bibitem{moore} C. Karrasch, R. Ilan and J.E. Moore, 
Phys. Rev. B{\bf 88}, 195129 (2013).
\bibitem{viti}
A. De Luca, J. Viti, L. Mazza, and D. Rossini, Phys.
Rev. B {\bf 90}, 161101 (2014).
\bibitem{kohn}
W. Kohn, Phys. Rev. {\bf 133}, 171 􏰛(1964).
\bibitem{fk} S. Fujimoto  and N. Kawakami, J. Phys. A. {\bf 31} 465 (1998).
\bibitem{psaroudaki} C. Psaroudaki and X. Zotos, to appear in 
J. Stat. Mech. (2016); arXiv:1502.05557.

\end{thebibliography}
\end{document}